\documentclass[conference]{IEEEtran}
\usepackage{cite}

\ifCLASSINFOpdf
   \usepackage[pdftex]{graphicx}
\else
   \usepackage[dvips]{graphicx}
\fi
\usepackage{amsmath}
\usepackage{array}
\usepackage{url}

\usepackage{comment}

\begin{document}
%
\title{Understanding the Heterogeneity of Contributors in Bug Bounty Programs}

%
\author{\IEEEauthorblockN{Hideaki Hata\IEEEauthorrefmark{1},
Mingyu Guo\IEEEauthorrefmark{2},
M. Ali Babar\IEEEauthorrefmark{2}}
\IEEEauthorblockA{\IEEEauthorrefmark{1}Graduate School of Information Science,
Nara Institute of Science and Technology, Japan\\ hata@is.naist.jp}
\IEEEauthorblockA{\IEEEauthorrefmark{2}School of Computer Science,
University of Adelaide, Australia\\ \{mingyu.guo, ali.babar\}@adelaide.edu.au}}


\maketitle

\begin{abstract}
\textit{Background:} While bug bounty programs are not new in software development, an increasing number of companies, as well as open source projects, rely on external parties to perform the security assessment of their software for reward. However, there is relatively little empirical knowledge about the characteristics of bug bounty program contributors.
\textit{Aim:} This paper aims to understand those contributors by highlighting the heterogeneity among them.
\textit{Method:} We analyzed the histories of 82 bug bounty programs and 2,504 distinct bug bounty contributors, and conducted a quantitative and qualitative survey.
\textit{Results:} We found that there are project-specific and non-specific contributors who have different motivations for contributing to the products and organizations.
\textit{Conclusions:} Our findings provide insights to make bug bounty programs better and for further studies of new software development roles.
\end{abstract}

%
\IEEEpeerreviewmaketitle

\section{Introduction}
Software vulnerabilities have significant impacts on software development and release management.
For example, before releasing \textsf{Firefox} 40 on August 11, \textsf{Mozilla} had to release security updates 39.0.3 on August 6,
2015, as an exploit was reported just before the planned release\cite{mozill2015blog}.
From the study of the National Vulnerability Database, Homaei and Shahriari reported that 
buffer errors, XSS, and access control problems were most reported vulnerabilities during studied seven years\cite{7854110}.
Bilge and Dumitras reported that after vulnerabilities are disclosed publicly, the volume of attacks
exploiting them increases by up to five orders of magnitude\cite{Bilge:2012:BWK:2382196.2382284}.
Given a general lack of available security experts and time to have software vulnerabilities internally assessed,
more and more software managers are opting for crowdsourcing solutions such as bug bounty programs.

A bug bounty program is a reward program offered by an organization to external parties, authorizing
them to perform security assessments on the organization's assets\cite{bugcrowd2016}.
In general, only the first report of a valid, that is, reproducible and fixable, vulnerability is rewarded;
others are considered duplicates and are not rewarded.
Vulnerabilities that cause a privileged escalation on the platform from an unprivileged to admin or administrator
are considered critical.
Vulnerabilities that severely affect multiple users or affect the security of the underlying platform 
are considered high-priority vulnerabilities. More significant vulnerabilities are rewarded with higher values.

\begin{table*}[t]
\centering
\caption{Comparison of crowdsourcing models. The contents of the bug bounty programs are presented by the authors and other contents were previously presented by LaToza and van der Hoek \cite{7367992}.}
\label{tab:comparison}
\begin{tabular}{l|ccc|c}
\hline
 & Peer production & Competitions & Microtasking & Bug Bounty \\
Dimension & (Open source) & (TopCoder) & (UserTesting.com) & Programs \\
\hline
Crowd size (necessary crowd size for tasks) & Small to medium & Small & Medium & Small \\
Task length (amount of time) & Hours to days & Week & Minutes & Hours to days \\
Expertise demands (required domain familiarity) & Moderate & Extensive & Minimal & Extensive \\
Locus of control (ownership of the creation) & Worker & Client & Client & Client \\
Incentives (factors motivating workers) & Intrinsic & Extrinsic & Extrinsic & Extrinsic \\
Task interdependence & Medium & Low & Low & Medium \\
Task context (amount of system info needed to know) & Extensive & Minimal & None & Exftensive \\
Replication (same task might be completed) & None & Several & Many & None \\
\hline
\end{tabular}
\end{table*}

LaToza and van der Hoek presented three factors that distinguish crowdsourcing from other outsourced
work: (1) the work is solicited through an open call to which basically anyone can respond, (2) the workers who
volunteer are unknown to the organization needing the work done, and (3) the group of workers can be
large\cite{7367992}. The authors compared different types of crowdsourcing work based on eight dimensions as shown
in Table \ref{tab:comparison}.
Since the workers in bug bounty programs need extensive domain expertise and extrinsic incentives, the programs
are closed to \textit{competition}-type crowdsourcing. In addition, bug bounty programs require extensive system information and
replication is not acceptable; these are the same characteristics of \textit{peer-production}-type crowdsourcing.
In sum, a bug bounty program can be considered a competitive crowdsourcing work with the same nature as
peer production.

Several previous empirical studies have helped to better understand bug bounty programs.
Munaiah and Meneely analyzed the Common Vulnerability Scoring System (CVSS) scores and bounty awarded for 703
vulnerabilities across 24 products, then found a weak correlation between CVSS scores and bounties\cite{Munaiah:2016:VSS:2989238.2989239}.
Finifer et al. analyzed datasets collected from \textsf{Chromium} and \textsf{Mozilla}, and reported that both bug bounty programs are economically efficient, compared to the cost of hiring full-time security researchers\cite{Finifter:2013:ESV:2534766.2534790}.
Zhao et al. studied \textsf{Wooyun} and \textsf{HackerOne}, bug bounty program platforms, and reported a significant
positive correlation between the expected bounty values and the number of received
vulnerabilities\cite{Zhao:2015:ESW:2810103.2813704}. They also reported that a considerable number of programs
showed a decreasing trend in vulnerability reports.
Hence, it is important that software managers gain a good understanding of bug bounty program contributors in order to build better programs instead of just increasing bounty values.

This study focuses on understanding the characteristics of bug bounty program contributors by
highlighting the heterogeneity of contributors to address the following research question:
\textit{What contributor heterogeneity exists in bug bounty programs?}
A deep understanding of bug bounty contributors will
help managers to reach potential contributors and to incentivize them effectively.
To answer this question, we examined past contribution histories in multiple bug bounty programs and conducted a
quantitative and qualitative survey, and then found that there are project-specific and non-specific contributors
who have different activities and motivations.

\section{Data Collection of Histories}

To avoid the bias in program selection, 
we made use of the following two publicly available lists of bug bounty programs:
a list provided by \textsf{bugsheet} in \url{http://bugsheet.com/directory/},
and a list provided by \textsf{Bugcrowd}, a company of bug bounty program platforms, in \url{https://bugcrowd.com/list-of-bug-bounty-programs/}.

\subsection{Selecting Programs}

In the above lists, some programs offer swags or gifts, or only publish contributors on their acknowledgements pages.
To focus on bounty programs providing monetary rewards, we ignored such programs.
Since bug bounty programs are a new trend and
retaining contributors is not easy\cite{Zhao:2015:ESW:2810103.2813704},
some programs disappeared when we accessed them.
If programs did not make contributor information public, we also ignored such programs.
We investigated these programs until early in August 2015, and found 82 active bug bounty programs
that provided \textit{Hall of Fame} for contributors;
33 programs were served in their own websites, and 49 programs were presented in bug bounty program
platforms, \textsf{Bugcrowd} or \textsf{HackerOne}. One program, \textsf{Barracuda}, had its own website,
but then moved to the \textsf{Bugcrowd} platform. We analyzed both websites to gather the data of contributors'
activities of \textsf{Barracuda}.

\begin{table}[t]
\centering
\caption{A Brief Summary of Studied Bug Bounty Programs, Approximate Bounty Ranges and Selected Programs.}
\label{tab:prog}
\begin{tabular}{llc}
\hline
Targets & Programs & Bounties \\
\hline
& Chromium & \\
Browser, OS, etc. & Microsoft & \$500 - \$100,000 \\
& Mozilla & \\
\hline
& AT\&T & \\
Web and/or Mobile & Facebook & \$25 - \$15,000 \\
& Yahoo & \\
\hline
& Coinbase & \\
Digital currency platform & Ethereum & 0.2 - 100 BTC \\
& Pikapay & \\
\hline
& PHP & \\
Programming Language & Python & \$50 - \$1,500 \\
& Ruby & \\
\hline
\end{tabular}
\end{table}

Table \ref{tab:prog} presents a brief summary of the obtained programs with their targets, selected programs, and summarized bounty ranges.
We selected a wide variety of bug bounty programs including
the popular browser and OS-related programs of \textsf{Chromium}, \textsf{Microsoft}, and \textsf{Mozilla}.
Programs targeting Web and/or mobile applications make up the majority of our list. In addition to major companies like \textsf{AT\&T}, \textsf{Facebook}, and \textsf{Yahoo}, there exist programs provided by startup companies and OSS projects.
Digital currency-related programs are another popular target class. There are nine such programs, and most of them prepared BTC bounties.
Some programs including \textsf{Flash}, \textsf{PHP}, \textsf{Python}, \textsf{Ruby}, etc. are provided by The Internet Bug Bounty sponsored by \textsf{Facebook}, \textsf{Microsoft}, and \textsf{HackerOne}.
There are programs provided by companies in Russia and the Netherlands.
If program descriptions and contributor parts were not written in English, we used
a translation service to read appropriate information.
However, we extracted the contributors' names or identifiers as they appeared without translation into English.

When focusing on bounties, browser and OS-related programs prepared higher bounties (\$500 - \$100,000).
Compared to these programs, Web and mobile application-related programs will pay lower bounties (\$25 - \$15,000).
When we consider 1 BTC as \$500.00,
the bounties of digital currency-related programs are relatively high values (about \$100.00 - \$50,000.00). \\

\subsection{Collecting Contributor Information}

For all 82 programs, we accessed their program websites and extracted contributor information.
We collected account names,
(full) names, and URLs (own websites, \textsf{Twitter} pages, \textsf{Facebook} pages,
\textsf{LinkedIn} pages, etc.) if provided
and the number of bounties
(reports) they contributed and rewarded.
Private accounts in \textsf{Bugcrowd} were ignored since we could not distinguish each account.
Although many programs ranked contributors (not based only on the number of reports, but also on their
severities), we did not extract rank information. If programs provided a contributor name
for each bounty, we counted the number of bounties for each contributor. Some programs only
showed contributor names in the whole period or for some periods (such as in years). In these cases, we considered
the number of names in all periods as the number of their contributions.

\begin{figure*}[t]
\centering
\includegraphics[width=0.95\linewidth]{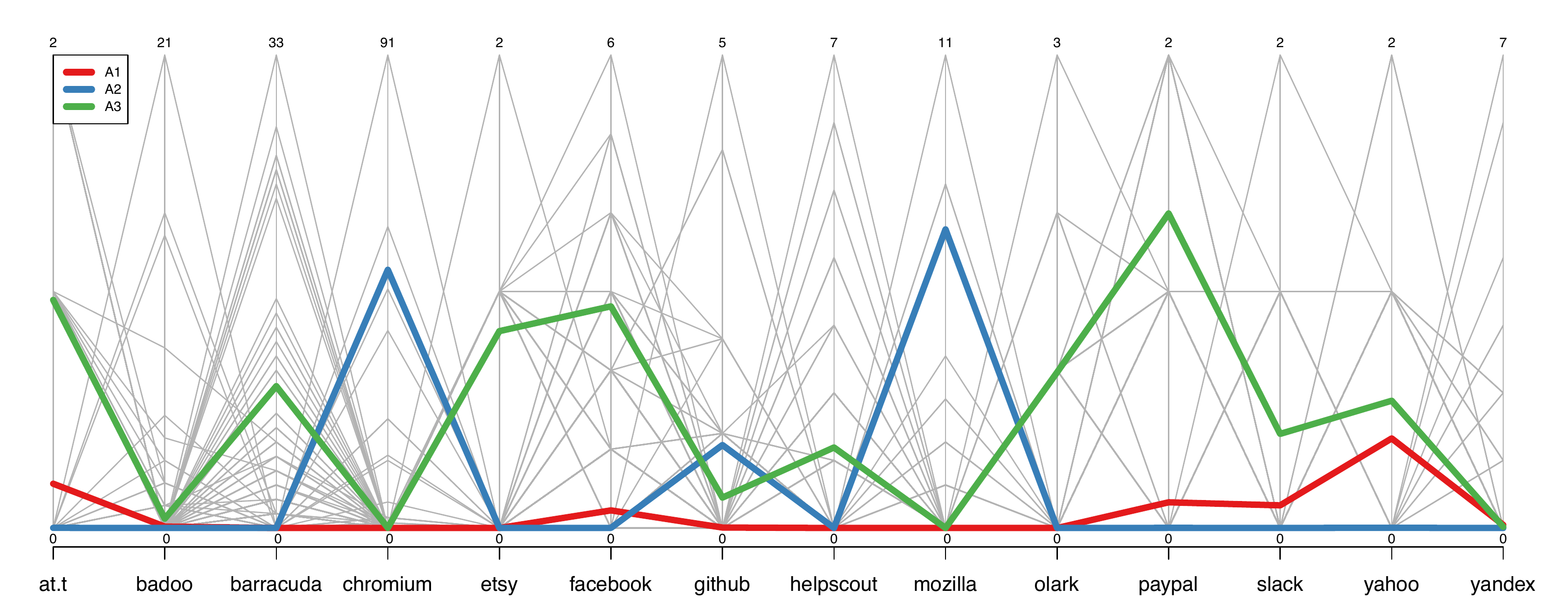}
\caption{Parallel coordinates of contributors. Colored lines represent three archetypes, A1, A2, and A3.}
\label{fig:aa_bar}
\end{figure*}

\subsection{Merging Contributor Identities}

From the collecting phase, we had more than 6,400 contributor names/accounts including duplicates. 
Similar to identity merging\cite{7816480},
we merged the same contributors' different names/accounts into single identifiers.
From the obtained contributor information, we considered a set of $<$ name, account name, websites $>$
as contributor identifiers. Depending on the data richness of program websites, some items could not be obtained
and some items had much information (there could be several URLs).
As shown in the report of \textsf{Bugcrowd} \cite{bugcrowd2016}, there were various contributors from different
countries, and some names were written in their own languages.
In addition, several contributors used special characters and emoji or emoticons for their identifiers;
many contributors showed their names or accounts inconsistently in different programs, and there were many
similar names on the list.
Because of these issues, we believe that preparing an accurate merging automation technique would be costly.
Therefore, we manually investigated all identifiers using sorting and searching terms repeatedly.
Some contributors showed full names for some programs, but shortened names (for example, without middle
names) for other programs. We considered them identical if common parts were the same.
\textsf{Twitter} accounts provided good clues because the same identifiers tend to be used for their account names.
In this merging process, we ignored contributor teams based on their names, for example, companies
and laboratories, because their activities are different from individual contributors' activities.
As a result, we obtained 2,504 distinct contributors.

\section{History Analysis}

\subsection{Demographics}

About $60\%$ of contributors have been rewarded only one bounty, and $67\%$ have contributed
to single program. Among 2,504 contributors in various programs, one-third of contributors
has contributed to multiple programs. So contributing to multiple programs
is not rare but common activities for bug bounty contributors.
Although top $20\%$ (500 contributors) contributed to $64\%$ bounties, the amount of bounties
worked by other less active contributors is also large, $36\%$. This implies that attracting
new contributors is important for bug bounty programs.

\subsection{Archetypal Analysis}

To analyze the characteristics of all contributors, we adopt an archetypal analysis.
An archetypal analysis is a statistical method that synthesizes a set of multivariate observations
through a few, not necessarily observed points (\textit{\textbf{archetypes}}),
which lie on the boundary of the
data scatter and represent pure individual types\cite{Porzio:2008:UAB:1416593.1416598}.
Archetypal analysis describes individual data points based on the distance from extreme points,
whereas cluster analysis focuses on describing its segments using the average members as the prototypes.
Marketing research is one of fields that have adopted archetypal analysis because it can provide well-separated
typical consumers\cite{desposito2006sa}, and therefore, we consider such analysis as also preferable for our study.
The \textsf{R} package \textsf{archetypes}\cite{Eugster:Leisch:2009:JSSOBK:v30i08} was used for this analysis.

To characterize contributors by the combinations of different programs in which they got bounties,
we prepared
an $n \times m$ matrix, which represents a multivariate dataset with $n$ contributors who had worked on
multiple programs (826) and $m$ programs that had more than 50 contributors who had more than one
bounty (14). Each value is the number of contributions for a program.
From the ``elbow criterion'' with the curve of the residual sum of squares (RSS), $k = 3$ is determined
as the number of archetypes.

Figure \ref{fig:aa_bar} presents a parallel coordinate plot of all bug bounty contributors.
Each line shows the number of contributions to different programs by one contributor.
The three colored lines are archetypes in this data (red is Archetype 1, blue is Archetype 2, and green is Archetype 3).
Note that the archetypes are not always actual contributors who are being observed, but are generated from the multidimensional data to be representatives of pure archetypes.
Then each observed contributor can be regarded as a mixture of those archetypes.
Obtained archetypes can be summarized as follows.

\begin{figure}[t]
\centering
\includegraphics[width=0.75\linewidth]{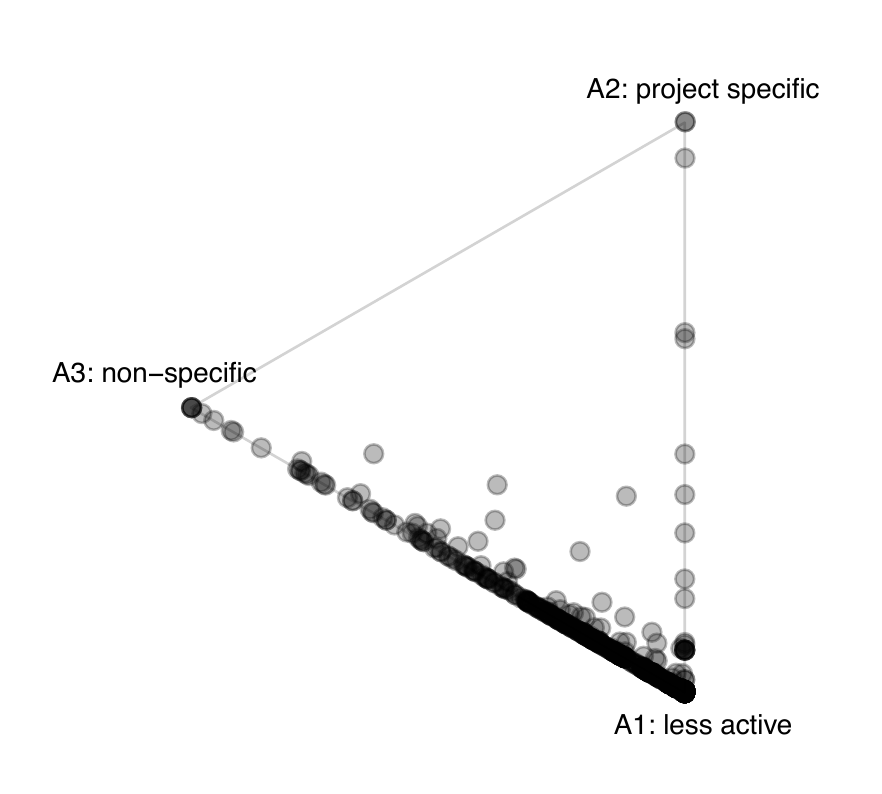}
\caption{A simplex visualization. Each dot represents a contributor, and the corners represent three archetypes A1, A2, and A3 as shown in Figure \ref{fig:aa_bar}. If a dot is close to a corner, it means that the contributor the dot represents is similar to the archetype.}
\label{fig:aa_map}
\end{figure}

\begin{itemize}
\item \textit{Archetype 1} (A1) is a less active contributor. Contributors close to this archetype might
have been rewarded a few bounties. They are likely to contribute to either \textsf{Yahoo}, \textsf{AT\&T}, \textsf{Paypal}, \textsf{Slack},
or \textsf{Facebook}.
\item \textit{Archetype 2} (A2) contributes only to the specific programs \textsf{Chromium} and \textsf{Mozilla},
and sometimes \textsf{GitHub}.
\item \textit{Archetype 3} (A3) contributes to many programs excluding \textsf{Chromium} and \textsf{Mozilla}.
\end{itemize}

Figure \ref{fig:aa_map} shows a simplex visualization\cite{simplexplot} of the results.
Each point, a contributor, is plotted among three archetypes.
A cluster near the line between A1 and A3 represents contributors who contributed less actively to actively for
various (not specific) programs, and most contributors belong to this cluster.
Although there are not so many contributors, there is another group of contributors near the line between
A1 and A2. These contributors tended to contribute to \textsf{Chromium} and \textsf{Mozilla}.
There are some contributors located between the two clusters.

\textbf{Observation 1}: \textit{Most contributors are less active (have only a few contributions).
There are two types of active contributors, project-specific and non-specific contributors,
and the latter is the majority of active contributors.}

Considering this observation, should managers pay attention to the different types of contributors? In order to answer
this question, we directly contacted both types of contributors.

\section{Quantitative and Qualitative Survey}
\label{sec:surv}

\def\mybar#1{
  #1 & {\rule{#1pc}{5pt}}}
\def\amybar#1{
  $#1\ast$ & {\rule{#1pc}{5pt}}}
\def\bmybar#1{
  $#1\dagger$ & {\rule{#1pc}{5pt}}}
\def\cmybar#1{
  $#1\ddagger$ & {\rule{#1pc}{5pt}}}
\def\dmybar#1{
  $#1\circ$ & {\rule{#1pc}{5pt}}}

\begin{table*}[t]
\footnotesize
\centering
\caption{Survey results from three A2 and four A3 contributors.}
\label{tab:survey}
\begin{tabular}{ll|rl|rl}
\multicolumn{2}{c}{Questions} & \multicolumn{2}{c}{A2 (Project-specific)} & \multicolumn{2}{c}{A3 (Non-specific)} \\
\hline
\multicolumn{6}{l}{Q1. What best describes your profile?} \\
\hline
& Security researcher & \mybar{3} & \mybar{1} \\
& Bug bounty hunter & \mybar{0} & \mybar{2} \\
& Developer & \mybar{1} & \mybar{0}\\
\hline
\multicolumn{6}{l}{Q2. What motivates you to report security bugs? (Multiple answers possible)} \\
\hline
& Making the products secure & \mybar{3} & \mybar{1} \\
& Helping users and developers & \mybar{3} & \mybar{1} \\
& Bounties & \mybar{3} & \mybar{3} \\
& Reputation & \mybar{3} & \mybar{2} \\
& Other & \amybar{1} & \mybar{0} \\
\hline
\multicolumn{6}{l}{Q3. How much time does it take on average to find and report a bug?} \\
\hline
& Less than 2 hours & \mybar{0} & \mybar{1} \\
& 2 to 6 hours & \mybar{0} & \mybar{2} \\
& 6 to 24 hours (1 day) & \mybar{0} & \mybar{0} \\
& 1 to 3 days & \mybar{1} & \mybar{0} \\
& 3 to 7 days (1 week) & \mybar{2} & \mybar{0} \\
& More & \mybar{0} & \mybar{0} \\
& Other & \bmybar{1} & \mybar{0} \\
\hline
\multicolumn{6}{l}{Q4. Why do you contribute to bug bounty programs? (Multiple answers possible)} \\
\hline
& I like the company/organization.
& \mybar{3} & \mybar{0} \\
& I use the products.
& \mybar{4} & \mybar{0} \\
& Part of my work is related to the products.
& \mybar{2} & \mybar{0} \\
& The products have many users.
& \mybar{2} & \mybar{1} \\
& Because of the amount of bounties.
& \mybar{2} & \mybar{2} \\
& I just found security bugs in the products.
& \mybar{2} & \mybar{2} \\
& It is open source.
& \mybar{2} & \mybar{0} \\
& The bug bounty program is well documented, prepared, and managed.
& \mybar{2} & \mybar{1} \\
& I am familiar with some bugs related to the product.
& \mybar{0} & \mybar{0} \\
& Other
& \mybar{0} & \cmybar{1} \\
\hline
\multicolumn{6}{l}{Q5. Do you have options to sell security bugs in black/gray markets?} \\
\hline
& Yes, if someone will buy them with high value. & \mybar{1} & \mybar{0} \\
& No. I will sell them to the vendors. & \mybar{2} & \mybar{2} \\
& No idea. & \mybar{0} & \mybar{1} \\
& Other & \dmybar{1} & \mybar{0} \\
\hline
\multicolumn{6}{l}{$\ast$ Contributing to the security of widely used open source programs.} \\
\multicolumn{6}{l}{$\dagger$ A few hours of my time, but days or weeks of computer time.} \\
\multicolumn{6}{l}{$\ddagger$ Gaining more experience dealing with new apps and technologies.} \\
\multicolumn{6}{l}{$\circ$ Receiving a bounty sum is a factor, but I think it is more important that I feel that my work is appreciated by the vendor.} \\
\end{tabular}
\end{table*}

\subsection{Selecting Contributors}

We selected all active contributors close to \textbf{A2 (project-specific)} and \textbf{A3 (non-specific)} archetypes,
and then searched their names in the Web.
If their email addresses were publicly available, we distributed a survey. We found nine email addresses,
and received three answers from A2 contributors and four answers from A3 contributors.
In general, it is difficult to obtain the contact information of bug bounty contributors because they tend not to make their profiles publicly available.
This implies that those bug bounty contributors are different from developers, who make their profiles open to the public.

\subsection{Quantitative Results}

Table \ref{tab:survey} summarizes the results of five questions. 
Regarding the profiles (Q1), A3 (non-specific) contributors considered themselves bug bounty hunters or a security researcher. However, A2 (project-specific) contributors did not consider themselves bug bounty hunters, but rather, security researchers or developers.
For A3 (non-specific) contributors, motivations for reporting security bugs (Q2) were mostly bounties. However, A2 (project-specific) contributors had various motivations; making the products secure, helping users and developers, and receiving bounties and reputation.

We found a significant difference in the time spent to find and report a bug (Q3). A3 (non-specific) contributors spent at most half a day, while A2 (project-specific) contributors tended to work from several days to one week on average. Since A3 (non-specific) contributors work on many programs, they could avoid work that required much time, or they could try their techniques or apply their knowledge to many programs using only about a half day's effort. However, A2 (project-specific) contributors focused on specific programs and spent much time to find and report security bugs. Our contributor categorization and the survey revealed that there are clearly different approaches in addressing bug bounty programs.

Question Q4 to contributors was: \textit{Why do you contribute to bug bounty programs?}
The majority of A3 (non-specific) contributors' answers were:
\textit{Because of the amount of bounties}, and
\textit{I found security bugs in the products}.
For A2 (project-specific) contributors, the majority of answers were:
\textit{I use the products}, and
\textit{I like the company/organization}.
Again we see that A2 contributors chose programs
because of their organizations and products.
These results confirm the findings from quantitative analysis, that is, why some contribute to specific programs and some contribute to many programs.

Finally, we asked a controversial question (Q5): \textit{Do you have options to sell security bugs in black/gray markets?}
Black markets are criminal interactions, and gray markets are interactions between sellers and government agencies,
as legal business deals\cite{fidler2015trade}. No A3 (non-specific) contributor reported the option to involve black or gray markets. However, one A2 (project-specific) contributor answered that s/he had the option of selling security bugs in those markets. Another A2 contributor selected \textit{Other} and left a comment that the bounty sum was a factor, but it was more important to feel that her/his work was appreciated by the vendor.
This might be reasonable because A3 contributors mainly work in many bug bounty programs so that they had less motivation to sell bugs to black/gray markets. However, A2 contributors expend much effort working on specific programs.
In other words, if they receive an unreasonable amount of bounties, some A2 contributors might think of selling bugs to
black/gray markets.
This implies that it is important for the bug bounty program organizers to consider the amount of effort required in order to attract contributors, especially by project-specific contributors.

\textbf{Observation 2}: \textit{Project-specific contributors work on programs because of the organizations and products, and tend to spend several days finding and reporting a bug, although non-specific contributors spend at most half a day on a project.}

As shown and discussed in Table \ref{tab:comparison}, a bug bounty program can be considered a competitive crowdsourcing
work with the nature of peer production. Project-specific contributors seem to have the mind of peer production workers.
As seen in the comment of an A2 (project-specific) contributor, appreciation for their work can be key in incentivizing a contributor to continue finding security bugs.

\subsection{Qualitative Results}

To uncover more details, we again contacted the survey respondents and had answers from two A2 (project-specific) and one A3 (non-specific) contributors.
In the previous survey, we did not distinguish human and machine time in Q3.
The A2 contributors revealed that they spend several human hours per week, but the computers run all the time.
Although A3 contributors mainly work in bug bounty program platforms, the A2 contributors have different ideas. One considers such platforms \textit{generally negative because any party other than the development project upstream is a potential security issue in itself}. Another contributor did not use platforms, but only reported vulnerabilities directly to the vendors. One A2 contributor considered all third-party programs as Gray market, and s/he has options to sell security bugs in Gray markets \textit{in theory}. Black markets were out of consideration for all three respondents.

What are bug bounty contributors expecting from bug bounty programs? Although the A3 (non-specific) contributor claimed a decent amount of bounties,
an A2 (project-specific) contributor required information flow in both directions. S/he thought it \textit{important to get information and stories from
the vendor's side, rather than just throw reports and get money in return}.
We think this an important clue for how bug bounty programs work with project-specific contributors.

\section{Discussion}

\subsection{Threats to Validity}

\textbf{Problems in merging contributor identifiers.}
Since this process has been conducted manually, there can be biases and mistakes in merging.
However, compared to automated techniques, we believe our manual inspections were more accurate.

\textbf{Limited datasets.}
Although we investigated all bug bounty programs in two separately maintained lists,
there can be other programs that have not appeared in these lists. This concern is related to external validity.
Although it is difficult to generalize our findings from our limited datasets, we would like to emphasize that our datasets contained various programs.

\textbf{Limited survey.}
Because of the difficulty of obtaining the contact information of the contributors, we have answers from only seven
bug bounty contributors. Further work should be beneficial to widen the scope of this survey. 

\subsection{Beyond Onion Models}

The onion models have been widely studied for open source software development communities
\cite{Nakakoji:2002:EPO:512035.512055,Ye:2003:TUM:776816.776867}.
``Onion'' refers to the successive layers of member types from core members, contributing developers,
bug reporters, and users, for example.
In the onion models, advancement through the member types is reward and recognition for each member's
abilities and achievements\cite{Aberdour:2007:AQO:1262538.1262594}, and developer initiation in OSS
depends on the social and technical actions of project
contributors\cite{Gharehyazie:2015:DIS:2821954.2821960}. Therefore, newcomers must first become
familiar with the code base, architecture, build environment, and work practices, which might take days
or weeks\cite{7367992}.
Bug bounty program contributors are almost outside of traditional software development hierarchies and onion models.
However, talented contributors are highly required and have a significant impact on software development.
Some studies have started analyzing monetary activities and impacts on software development ecosystems\cite{7884685,7925419}.

\section{Conclusion}
Since security bugs can have significant impact on software development and users, attracting and managing bug bounty contributors present new challenges.
This study revealed that there are two different types of contributors.
Most bug bounty contributors are non-project-specific bug bounty hunters and are different from traditional contributors who work on specific projects.
Although there are not many, but also there exist project-specific security contributors.

Our practical recommendation for program managers is that they first need to know whether
there are project-specific specialists in their software development communities,
and then they need to think of retaining them by expressing appreciation and preparing channels to communicate, for example.
Moreover, program managers also need to consider attracting non-specific security specialists with reasonable bounties.


%



\section*{Acknowledgment}
This work has been supported by JSPS KAKENHI Grant Number 16H05857 and JSPS Program for Advancing Strategic International Networks to Accelerate the Circulation of Talented Researchers: Interdisciplinary Global Networks for Accelerating Theory and Practice in Software Ecosystem (G2603).


\ifCLASSOPTIONcaptionsoff
  \newpage
\fi


\begin{thebibliography}{10}
\providecommand{\url}[1]{#1}
\csname url@samestyle\endcsname
\providecommand{\newblock}{\relax}
\providecommand{\bibinfo}[2]{#2}
\providecommand{\BIBentrySTDinterwordspacing}{\spaceskip=0pt\relax}
\providecommand{\BIBentryALTinterwordstretchfactor}{4}
\providecommand{\BIBentryALTinterwordspacing}{\spaceskip=\fontdimen2\font plus
\BIBentryALTinterwordstretchfactor\fontdimen3\font minus
  \fontdimen4\font\relax}
\providecommand{\BIBforeignlanguage}[2]{{%
\expandafter\ifx\csname l@#1\endcsname\relax
\typeout{** WARNING: IEEEtran.bst: No hyphenation pattern has been}%
\typeout{** loaded for the language `#1'. Using the pattern for}%
\typeout{** the default language instead.}%
\else
\language=\csname l@#1\endcsname
\fi
#2}}
\providecommand{\BIBdecl}{\relax}
\BIBdecl

\bibitem{mozill2015blog}
\BIBentryALTinterwordspacing
D.~Veditz. (2015) Firefox exploit found in the wild.
  \url{ttps://blog.mozilla.org/security/2015/08/06/firefox-exploit-found-in-the-wild/}
\BIBentrySTDinterwordspacing

\bibitem{7854110}
H.~Homaei and H.~R. Shahriari, ``Seven years of software vulnerabilities: The
  ebb and flow,'' \emph{{IEEE} Security Privacy}, vol.~15, no.~1, pp. 58--65,
  Jan 2017.

\bibitem{Bilge:2012:BWK:2382196.2382284}
\BIBentryALTinterwordspacing
L.~Bilge and T.~Dumitras, ``Before we knew it: An empirical study of zero-day
  attacks in the real world,'' in \emph{Proc.\ of 2012 {ACM} Conf. on Computer
  and Communications Security}, ser. CCS '12.\hskip 1em plus 0.5em minus
  0.4em\relax New York, NY, USA: ACM, 2012, pp. 833--844.
\BIBentrySTDinterwordspacing

\bibitem{bugcrowd2016}
\BIBentryALTinterwordspacing
{Bugcrowd INC.}, ``2016 state of bug bounty report,'' 2016.
  \url{https://pages.bugcrowd.com/2016-state-of-bug-bounty-report}
\BIBentrySTDinterwordspacing

\bibitem{7367992}
T.~LaToza and A.~van~der Hoek, ``Crowdsourcing in software engineering: Models,
  motivations, and challenges,'' \emph{{IEEE} Softw.}, vol.~33, no.~1, pp.
  74--80, Jan 2016.

\bibitem{Munaiah:2016:VSS:2989238.2989239}
\BIBentryALTinterwordspacing
N.~Munaiah and A.~Meneely, ``Vulnerability severity scoring and bounties: Why
  the disconnect?'' in \emph{Proc. of 2nd Int. Workshop on Soft. Analytics},
  ser. SWAN '16.\hskip 1em plus 0.5em minus 0.4em\relax New York, NY, USA: ACM,
  2016, pp. 8--14.
\BIBentrySTDinterwordspacing

\bibitem{Finifter:2013:ESV:2534766.2534790}
\BIBentryALTinterwordspacing
M.~Finifter, D.~Akhawe, and D.~Wagner, ``An empirical study of vulnerability
  rewards programs,'' in \emph{Proc.\ of 22nd {USENIX} Conf. on Security}, ser.
  SEC '13.\hskip 1em plus 0.5em minus 0.4em\relax Berkeley, CA, USA: USENIX
  Association, 2013, pp. 273--288.
\BIBentrySTDinterwordspacing

\bibitem{Zhao:2015:ESW:2810103.2813704}
\BIBentryALTinterwordspacing
M.~Zhao, J.~Grossklags, and P.~Liu, ``An empirical study of web vulnerability
  discovery ecosystems,'' in \emph{Proc.\ of 22nd {ACM} {SIGSAC} Conf. on
  Computer and Communications Security}, ser. CCS '15.\hskip 1em plus 0.5em
  minus 0.4em\relax New York, NY, USA: ACM, 2015, pp. 1105--1117.
\BIBentrySTDinterwordspacing

\bibitem{7816480}
I.~S. Wiese, J.~T. d.~Silva, I.~Steinmacher, C.~Treude, and M.~A. Gerosa, ``Who
  is who in the mailing list? comparing six disambiguation heuristics to
  identify multiple addresses of a participant,'' in \emph{Proc.\ of 32nd
  {IEEE} Int. Conf. on Softw. Maintenance and Evolution}, Oct 2016, pp.
  345--355.

\bibitem{Porzio:2008:UAB:1416593.1416598}
\BIBentryALTinterwordspacing
G.~C. Porzio, G.~Ragozini, and D.~Vistocco, ``On the use of archetypes as
  benchmarks,'' \emph{Appl. Stoch. Model. Bus. Ind.}, vol.~24, no.~5, pp.
  419--437, Sep. 2008.
\BIBentrySTDinterwordspacing

\bibitem{desposito2006sa}
M.~R. {D'Esposito}, F.~Palumbo, and G.~Ragozini, ``Archetypal analysis for
  interval data in marketing research,'' \emph{Italian J. Appl. Statist.},
  vol.~18, no.~2, 2006.

\bibitem{Eugster:Leisch:2009:JSSOBK:v30i08}
\BIBentryALTinterwordspacing
M.~J.~A. Eugster and F.~Leisch, ``From {Spider-Man} to hero --- archetypal
  analysis in {R},'' \emph{J. of Statistical Softw.}, vol.~30, no.~8, pp.
  1--23, 4 2009.
\BIBentrySTDinterwordspacing

\bibitem{simplexplot}
\BIBentryALTinterwordspacing
S.~Seth and M.~Eugster, ``\BIBforeignlanguage{English}{Probabilistic archetypal
  analysis},'' \emph{\BIBforeignlanguage{English}{Machine Learning}}, pp.
  1--29, 2015.
\BIBentrySTDinterwordspacing

\bibitem{fidler2015trade}
\BIBentryALTinterwordspacing
M.~Fidler, ``Regulating the zero-day trade: A preliminary analysis,''
  \emph{I/S: A J. of Law and Policy for the Inf. Society}, vol.~11, December
  2015.
\BIBentrySTDinterwordspacing

\bibitem{Nakakoji:2002:EPO:512035.512055}
\BIBentryALTinterwordspacing
K.~Nakakoji, Y.~Yamamoto, Y.~Nishinaka, K.~Kishida, and Y.~Ye, ``Evolution
  patterns of open-source software systems and communities,'' in \emph{Proc.\
  of 5th Int. Workshop on Principles of Softw. Evolution}, ser. IWPSE
  '02.\hskip 1em plus 0.5em minus 0.4em\relax New York, NY, USA: ACM, 2002, pp.
  76--85.
\BIBentrySTDinterwordspacing

\bibitem{Ye:2003:TUM:776816.776867}
\BIBentryALTinterwordspacing
Y.~Ye and K.~Kishida, ``Toward an understanding of the motivation open source
  software developers,'' in \emph{Proc.\ of 25th Int. Conf. on Softw. Eng.},
  ser. ICSE '03.\hskip 1em plus 0.5em minus 0.4em\relax Washington, DC, USA:
  IEEE Computer Society, 2003, pp. 419--429.
\BIBentrySTDinterwordspacing

\bibitem{Aberdour:2007:AQO:1262538.1262594}
\BIBentryALTinterwordspacing
M.~Aberdour, ``Achieving quality in open source software,'' \emph{{IEEE}
  Softw.}, vol.~24, no.~1, pp. 58--64, Jan. 2007.
\BIBentrySTDinterwordspacing

\bibitem{Gharehyazie:2015:DIS:2821954.2821960}
\BIBentryALTinterwordspacing
M.~Gharehyazie, D.~Posnett, B.~Vasilescu, and V.~Filkov, ``Developer initiation
  and social interactions in {OSS}: A case study of the apache software
  foundation,'' \emph{Empirical Softw. Eng.}, vol.~20, no.~5, pp. 1318--1353,
  Oct. 2015.
\BIBentrySTDinterwordspacing

\bibitem{7884685}
T.~Kanda, M.~Guo, H.~Hata, and K.~Matsumoto, ``Towards understanding an
  open-source bounty: Analysis of bountysource,'' in \emph{Proc.\ of 24th
  {IEEE} Int. Conf. on Softw. Analysis, Evolution and Reengineering}, ser.
  SANER '17, Feb 2017, pp. 577--578.

\bibitem{7925419}
K.~Nakasai, H.~Hata, S.~Onoue, and K.~Matsumoto, ``Analysis of donations in the
  eclipse project,'' in \emph{Proc.\ of 8th Int. Workshop on Empirical Softw.
  Eng. in Practice}, ser. IWESEP '17, March 2017, pp. 18--22.

\end{thebibliography}
\end{document}